# User Modeling Combining Access Logs, Page Content and Semantics

Blaž Fortuna
Jožef Stefan Institute
Jamova cesta 39
1000 Ljubljana, Slovenia
+386 (1) 477 3528

blaz.fortuna@ijs.si

Dunja Mladenić
Jožef Stefan Institute
Jamova cesta 39
1000 Ljubljana, Slovenia
+386 (1) 477 3528

dunja.mladenic@ijs.si

Marko Grobelnik
Jožef Stefan Institute
Jamova cesta 39
1000 Ljubljana, Slovenia
+386 (1) 477 3528

marko.grobelnik@ijs.si

## ABSTRACT
The paper proposes an approach to modeling users of large Web sites based on combining different data sources: access logs and content of the accessed pages are combined with semantic information about the Web pages, the users and the accesses of the users to the Web site. The proposed approach represents each user by a set of features derived from the different data sources, where some feature values may be missing for some users. It further enables user modeling based on the provided characteristics of the targeted user subset. The approach is evaluated on real-world data where we compare performance of the automatic assignment of a user to a predefined user segment when different data sources are used to represent the users.

## Categories and Subject Descriptors
H.3.5 [**Programming Languages**]: On-line Information Services – *Web-based services.*

## 1. INTRODUCTION
Large Web sites focused on providing content attract different kinds of users depending on the content and their reputation. Understanding of user populations can, among others, be used to provide better navigation and promotion of web site's assets and can give an important advantage when talking with the advertisers. To achieve this, some sites offer more value to the users for the price of free registration, where registration usually means acquiring additional demographic information about the user, such as, age, gender, education etc. They use this information to understand which user segments are visiting different sections of the web sites (e.g. sports or cooking section). However, the sparseness of the registration data can be prohibitive in providing deeper insights on the user population when focusing on smaller parts, individual articles and the context of a single user's visit (e.g. referring domain or time of the day).

In this paper we present an approach to building flexible, comprehensive and scalable user models which take into account three components: access logs, page content and the users' registration information. We refer to this additional information on users as semantic information that is available for only a small subset of the users. The sites focused on providing content usually provide some additional semantic information on the content of their Web pages, such as, author of the page content, topic category of the page content, named entities that are mentioned on a page etc. Access logs in addition to URL of the accessed page contain information on the user location, time, device, etc. This can be seen as semantic information on the access of the users to the web site.

The presented approach consists of the following steps. Firstly, the user segment is defined. Segments can be as broad as *"female users"* to narrower one, such as, *"female users older than 30 years that at least once visited a page from the style section category"*. The connection of the segment size with the quality of the user model is studied in more details in the experimental section. Secondly, a training set is assembled by identifying all the users from the defined segment, and complementing them with a random sample of the remaining users. A feature vector is constructed for each user, containing the information of all the pages they visited and the contexts in which they were visited. Finally, Support Vector Machine (SVM) [1] is used to build a linear model of the segment.

The main contribution of this paper are the first two steps in the proposed approach and, evaluation and stress testing on real-world data of a large web-site with 5 million unique users per day, a million of which are registered. Additionally, the paper describes how such approach can be implemented in a scalable way so it can be used on near real-time stream of access logs by users who are not experts in web mining.

The remaining of the paper is structured as follows. First we give an overview of related work, followed by the description of the data which goes into the models. Then we describe the proposed approach, a summary of the evaluation on a real-world dataset and finish with conclusions.

## 2. RELATED WORK
A number of methods for mining the Web have been applied in the last two decades including crawling, indexing and modeling the data using unsupervised, semi-supervised and supervised machine learning methods as well as, using social network analysis [13]. There are several areas in the web mining community [5] where content and access logs were used to profile users. There, the authors mostly focus on the application of profiling to recommender systems and e-commerce based on query logs. This paper focuses on user modeling for large web sites providing content, such as magazines or newspapers, with the task of classifying users into pre-defined demographic segments.

Semantic information providing domain knowledge has been used in web personalization mainly related to content or items in recommendation systems [16]. Incorporating some level of semantics in textual content analysis was proposed by several researchers (e.g., [17], [18]) showing a number of open challenges. In the proposed user modeling we also incorporate semantics of textual content including manually assigned document keywords, annotation of named entities occurring inside the page content and, topic categories assigned to each document based on a predefined topic ontology.

User modeling and web personalization, in addition to adding semantic information to web page content in access logs analysis (e.g., [14]) and web search (e.g., [15]), can also consider having some background information on the users obtained via questionnaires (e.g., [11]) i.e., semantic information on the users. In the area of semantic web mining [6] several approaches have been proposed to combine higher level annotation data (e.g., categories, tags) with the content and access logs as means of fighting the sparseness of training data. In addition, usage of ontologies to describe user interest [15] and behavior can bring semantics in web mining but is expected to require additional efforts for constructing appropriate ontologies. As pointed out in [12], the future of web mining to large extent depends on the development of semantic web.

Work has been also done in predicting demographic characteristics of users based on the pages they visited. For example, in [2] authors used the content and categories of the pages visited by the users to predict their gender and age group. They proposed combination of SVM and Latent Semantic Indexing [7] to reduce the sparseness in the data. User profiling was also prominent topic in ad targeting community with most of the work focusing on personalizing display ads process to maximize clicks [8][9][10]. In our work we do not directly address ad targeting, even though the proposed modeling of user segments can be exploited in ad targeting. Our approach differs from the related work in combining access logs and page content with semantic information from different data sources. Moreover, we enable user modeling based on characteristics of the targeted user subset, where some data may be missing for some of the users.

## 3. PROPOSED APPROACH

The data used in the modeling comes from three main sources: access logs, page content and the user registration data. In this section we provide details and breakdown of each of these sources as implemented in a real-world application of the proposed approach for a large international content provider.

Access logs record the interactions between the user and the web site. Basic element of the interaction is a request for a page on the web site by specifying its URL. Each interaction is also described by its context:

- *Time*: the time of the interaction,
- *Location*: decoding of the user's IP using GeoIP service[1],
- *Source*: referring page with the search terms when the user is coming from a search engine,
- *Device*: user agent string, specifying the browser, operating system and the device.

Access logs also store the list of cookies deposited on the user's side by the web site. This mechanism is used to uniquely identify users over more interactions by storing a unique ID the first time a user visits the site. The unique ID is also be used to link the registered users to their demographic profile.

Each interaction revolves around the content contained within the requested page. The page and its content can be described using low level features (e.g. keywords), annotations (e.g. named entities), aggregate features (e.g. topic categories), content metadata (e.g. author, publish date) and page metadata (e.g. page type). In the case of the studied web site all of the mentioned features were available but not all for each page. In the case of named entities, manually constructed list was extended with a list of automatically extracted named entities [19]. Topic categories, manually assigned by authors and/or web site editors, were also extended with automatic classification into DMoz [20].

---

[1] http://www.maxmind.com/app/locate_ip

Registration data was provided for all registered users. It contained, among others, fields describing the age, gender, income and the job title. Since the data was collected over the web, it is prone to some level of noise. Experiments in Section 5 provide more details on the robustness of the gathered data.

## 4. MODELING AND IMPLEMENTATION

The first step of the presented approach is the definition of a user segment. User segment is a subset of users, web site's visitors, which share a common characteristic. Simple examples of standard market segments are breakdown of users by gender or age [2]. More complex segmentation can involve income, job or industry, interests a combination of these. The context from the access logs provides additional segments that are relevant to a web site, such as users coming from search engines (e.g. Google, Bing) or social networks (e.g. Facebook, Twitter), users also reading about traveling, users from particular country or city, etc.

It is important to keep this step simple enough, to let the domain experts with little or no knowledge of web mining to specify and define the segments. To achieve this, our implementation stores all the visitors in an inverted index where they are searchable by their demographics data and visits. This changes the task of defining a user segment to a more familiar task of specifying a search query. Figure 1 shows the list of fields describing the users that are implemented in the system and Figure 2 shows a sample query that can be used to define user segment.

Second step of the presented approach is feature extraction and formation of training set from the defined user segment. The features describing each user visit are extracted from the fields shown in Figure 1. Vector space model [3], standard text representation from text mining, is used to represent the data. The vector elements coming from each of the fields are normalized so they contribute same overall weight to the final norm of the vector. This representation is used due to its flexibility and serves as a good fit for linear SVM model, due to its capability to handle over-fitting and missing elements (each element, e.g. referring domain, is just one of many elements in the linear score).

Training set consists of users considered as data points. Each user is either in a positive class, if he/she matches the segment definition, or in a negative class otherwise. Each user is described by a centroid of all its visits' feature vectors. Sampling can be used at this step to (a) limit the amount of users for which the feature vectors need to be prepared and (b) limit the overwhelming size of negative class in the case of small user segments. Sampling can be necessary due to the scale of traffic on larger web sites (10s of millions of users with 100s of millions of visits per month in the case of web site used in this paper). Also, the cost of training set extraction and model training can at some point outweigh the diminishing returns on the increase of performance.

The third and final step is applying linear SVM to the training set. Support vector machine is a family of algorithms that has gained a wide recognition as one of the state-of-the-art machine learning algorithms for tasks such as classification, regression, etc. In the basic formulation they try to separate two sets of training examples by hyperplane that maximizes the margin (distance between the hyperplane and the closest points). In addition one usually permits few training examples to be misclassified.

| Domain | Referring Search Term | Day of the Week |
|---|---|---|
| Sub-domain | Referring Domain | Hour of the day |
| Page URL | Referring URL | User Agent |
| Page Meta Tags | Country (from IP) | *Income* |
| Page Title | State (from IP) | *Age* |
| Page Content | City (from IP) | *Gender* |
| Named Entities | Date | |

**Figure 1. Fields which can be combined to specify a user segment. Emphasized fields are the ones extracted from the user registration data.**

**Gender** = *female*
**Income** ≥ *$100,000*
**Meta Data** = *Category Style*

**Figure 2. Example query used to specify a user segment. It selects all the users registered as female earning more than $100,000 per year which at least once visited a page from category Style.**

BOOK CANCER CHILDREN CHOP DESIGNED DR EAT FAMILY FOODS HAIR HOME HOUSE KENNEDY MS RESEARCH SCHOOLS STUDENTS STUDY WOMEN

**Figure 3. Tag cloud describing the segment from Figure 2.**

The trained model can be used in several ways. First and most obvious one is to apply the model to the unregistered users. However, trained SVM model can be visualized using feature selection approach specified in [4]. Keywords or features describing the user segments are the ones, whose weights in SVM normal vector contribute most when deciding if the centroid of the positive class is positive. Figure 3 shows top keywords describing the segment defined in Figure 2. This information was found in practice to be useful to the web site editors and marketers to better understand the behavior of their users (e.g. what does a particular segment of users typically read).

## 5. EXPERIMENTS

This section presents a summary of the experimentation performed on a real-world dataset obtained by parsing logs of a major news publishing web site. To evaluate the proposed approach we tested it against several demographic dimensions: gender, age and income. For each of these dimensions we tested the influence of different feature sets and the frequency of visits by users in the training set. Gender demographic dimension had two possible classes, male and female, with 250,000 examples. Table 1 lists the categories and number of examples other two dimensions.

**Table 1 List of categories along Age demographic dimension (left) and Income demographic dimension (right).**

| Category | Size | Category | Size |
|---|---|---|---|
| 21-30 | 100,000 | 0-24k | 50,000 |
| 31-40 | 100,000 | 25k-49k | 50,000 |
| 41-50 | 100,000 | 50k-74k | 50,000 |
| 51-60 | 100,000 | 75k-99k | 50,000 |
| 61-80 | 100,000 | 100k-149k | 50,000 |
| | | 150k-254k | 50,000 |

Classification results were measured using two metrics. First was Break Even Point (BEP) – a hypothetical point at which precision (ratio of positive documents among retrieved ones) and recall (ratio of retrieved positive documents among all positive documents) are the same. This measure does not depend on the threshold parameter returned by classification algorithm. The second measure is ROC curve, showing how number of true and false positive change as we decrease the threshold. These two measures were selected as they capture different aspects of the interaction between relevance and reach relative to the threshold position. In the case of ROC curve, the y-axis (true positive rate) shows the ratio of relevant users captured in the segment and the x-axis (false positive rate) shows the ratio of users incorrectly assigned to the segment. All presented results were obtained using SVM model and 5-fold cross-validation.

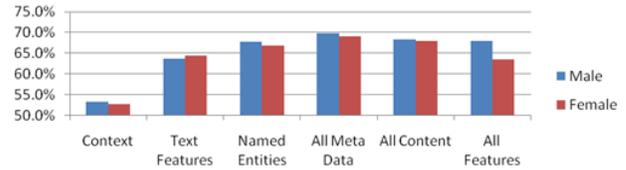

**Figure 4. BEP for Gender (50% = random)**

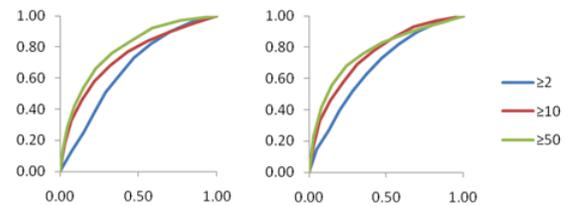

**Figure 5. ROC curve for Male (left) and Female (right).**

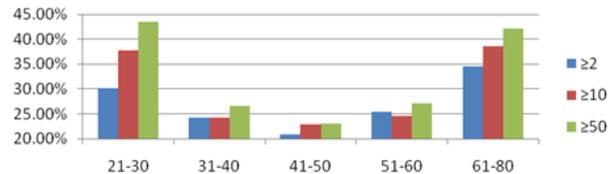

**Figure 6. BEP for Age (20% = random)**

For each demographic dimension two sets of experiments were performed. The first set evaluated classification performance of different feature sets used to describe a user:

- *Context* – features that can be obtained from access logs, such as time, location, source and device.
- Content features:
    - *Text Features* – keywords extracted from the articles read by the user.
    - *Named Entities* – named entities extracted using information extraction techniques [18]
    - *All Metadata* – metadata assigned to the article by the editors; e.g. byline, topics, main keywords, people, organization and countries mentioned in the article, date.
- *All Content* – combination of text features, named entities and metadata features.
- *All Features* – combination of all above features.

Figure 4 shows classification results for Gender demographic dimension for different combinations of feature sets. It can be seen, that metadata was best performing feature set for both Male and Female categories. It is interesting that combining all feature sets can actually significantly hurt performance in the case of Female class. Figure 5 shows ROC curve for Male and Female class for different minimal number of visits required by users. It can be see that, as expected, performance increases as the user history increases. Overall, it is interesting to see, that just by looking at metadata of articles, one can predict gender with 70% BEP.

Figure 6 shows more interesting results. When predicting an age group of the user, it turns out that younger (21-30) or older (61-80) users are significantly easier to predict compared to the middle age groups. Prediction results for the 41-50 category were only slightly above random. Named entities turned out to be most significant feature set. As expected, performance increases as more observations are available for a user.

Income category is very hard to predict based on what news articles people read. Nevertheless, it can be seen that, similar as in the case of age groups, the edge case (150-254) is easier to predict (~22% BEP) compared to more common categories (~16% BEP). Context features turned out to have significant contributions in the income classifications with geographic location extracted from IP address being the most informative feature.

## 6. CONCLUSIONS

The presented paper focuses on modeling the users with the purpose of understanding the demographics coming to various parts of a web-site in different contexts. To achieve this only content based profiling is not sufficient anymore. The proposed approach combines semantically enriched content information with context information from access logs such as referring pages, time, geographic location and device information.

Experiments show that different segmentations, such as gender, age or income, relay on different feature subsets for classification. For example, metadata turned out significant in the case of Gender segmentation compared to named entities for age and context, or more specifically geographic location, for income segmentations.

The advantages of the presented approach are two-fold. First, the proposed methodology allows the domain expert to specify complex user segments and through visualization understand the behavior of the segment. Second, it allows for building more accurate models which can be used to generalize the demographics to the users for which there is none available.

## 7. REFERENCES

[1] Cristianini N. & Shawe-Taylor, J., An introduction to support vector machines, Cambridge University Press

[2] Hu, J., Zeng, H., Li, H., Niu, C., and Chen, Z. 2007. Demographic prediction based on user's browsing behavior. In Proceedings of the 16th international Conference on World Wide Web (Banff, Alberta, Canada, May 08 - 12, 2007). WWW '07. ACM, New York, NY, 151-160.

[3] Salton, G. Developments in Automatic Text Retrieval. Science, Vol 253, 974-979

[4] Brank J., Grobelnik M., Milic-Frayling N. & Mladenic D. Feature selection using support vector machines. Proc. of the Third International Conference on Data Mining Methods and Databases for Engineering, Finance, and Other Fields, 2002.

[5] Nasraoui, O., Spiliopoulou, M., Zaïane, O. R., Srivastava, J., and Mobasher, B. 2008. WebKDD 2008: 10 years of knowledge discovery on the web post-workshop report. SIGKDD Explor. Newsl. 10, 2 (Dec. 2008), 78-83.

[6] B. Berendt, A. Hotho, and G. Stumme, Towards semantic web mining. Proceedings of the First International Semantic Web Conference, 2002 (Springer, Sardinia, 2002), pp. 264-278

[7] Deerwester S., Dumais S., Furnas G., Landuer T. & Harshman R. Indexing by Latent Semantic Analysis, J. of the American Society of Information Science, vol. 41/6, 391-407

[8] Ghosh, A., Rubinstein, B. I., Vassilvitskii, S., and Zinkevich, M. 2009. Adaptive bidding for display advertising. In Proceedings of the 18th international Conference on World Wide Web (Madrid, Spain, April 20 - 24, 2009). WWW '09. ACM, New York, NY, 251-260.

[9] Joanna Jaworska, Marcin Sydow. Behavioural Targeting in On-Line Advertising: An Empirical Study. Web Information Systems Engineering - WISE 2008

[10] Li, Y., Surendran, A. C., and Shen, D. 2007. Data mining and audience intelligence for advertising. SIGKDD Explor. Newsl. 9, 2 (Dec. 2007), 96-99.

[11] Eirinaki, M., Vazirgiannis, M. 2003. Web mining for web personalization, ACM Transactions Internet Technology, 3/1, pp. 1-27, ACM, New York, NY, USA.

[12] Berendt, B., Hotho, A., Mladenic, D., Someren, M. van, Spiliopoulou, M., Stumme, G. 2004. A roadmap for web mining : from web to semantic web. In Web mining : from web to semantic web, September, 2003 (Lecture notes in artificial intelligence, Lecture notes in computer science, vol. 3209), Berlin; Heidelberg; New York: Springer, pp. 1-22.

[13] Chakrabarti, S. 2002, Mining the Web: Discovering Knowledge from Hypertext Data, Morgan-Kauffman.

[14] Jorge, A., Alves, M. A., Grobelnik, M., Mladenic, D., Petrak, J. 2003. Web site access analysis for a national statistical agency. In Data mining and decision support: integration and collaboration, The Kluwer international series in engineering and computer science, SECS 745, Boston; Dordrecht; London: Kluwer Academic Publishers, pp. 167-176.

[15] Sieg, A., Mobasher, B., Burke, R. 2007. Ontological User Profiles for Representing Context in Web Search. Proceedings of the Workshop on Web Personalization and Recommender Systems. At the ACM International Conference on Web Intelligence, Silicon Valley, CA, November 2007.

[16] Singh Anand, S., Mobasher, B. 2005. Intelligent Techniques for Web Personalization. In Intelligent Techniques for Web Personalization, Lecture Notes in Artificial Intelligence (LNAI 3169), Springer, 2005.

[17] Baxter, D., Klimt, B., Grobelnik, M., Schneider, D. I., Witbrock, M. J., Mladenic, D. 2009. Capturing document semantics for ontology generation and document summarization. In Semantic knowledge management: integrating ontology management, knowledge discovery, and human language technology. Berlin; Heidelberg: Springer, 2009, pp. 141-154.

[18] Rusu, D., Fortuna, B., Grobelnik, M., Mladenic, D. 2009. Semantic graphs derived from triplets with application in document summarization. Informatica Journal, vol. 33, no. 3, pp. 357-362.

[19] Stajner, T., Mladenic, D. 2009. Entity Resolution in Texts Using Statistical Learning and Ontologies, In Proceedings of the fourth Asian Semantic Web Conference ASWC 2009.

[20] Grobelnik, M., Mladenic, D. 2005. Simple classification into large topic ontology of web documents. Journal of Computing and Information Technology, 2005, vol. 13, pp. 279-285